%% file: MAIN.tex
\documentclass[conference]{IEEEtran}
\IEEEoverridecommandlockouts

\usepackage{cite}
\usepackage{amsmath,amssymb,amsfonts}
\usepackage{graphicx}
\usepackage{textcomp}
\usepackage{xcolor}

\input{preamble}

\def\BibTeX{{\rm B\kern-.05em{\sc i\kern-.025em b}\kern-.08em
    T\kern-.1667em\lower.7ex\hbox{E}\kern-.125emX}}

\begin{document}

\title{MOTIF: A tool for Mutation Testing with Fuzzing

}

\input{authors-IEEE}

\maketitle

\input{sections/abstract}


\input{sections/introduction}

\input{sections/approach}
\input{sections/evaluation}
\input{sections/conclusion}

\section*{Acknowledgment}
This research was supported by ESA via a GSTP element contract (RFQ/3-17554/21/NL/AS/kkIMPROVE) and by the NSERC Discovery and Canada Research Chair programs.
Our experiments were performed using the HPC facilities of the University of Luxembourg (see http://hpc.uni.lu).

\bibliographystyle{IEEEtran}
\bibliography{bibliography/ref}

\end{document}

%% file: preamble.tex
\usepackage{amsmath}
\usepackage{array}
\usepackage{lipsum}
\usepackage{listings}
\usepackage{pmboxdraw}  

\definecolor{mGreen}{rgb}{0,0.6,0}
\definecolor{mGray}{rgb}{0.5,0.5,0.5}
\definecolor{mPurple}{rgb}{0.58,0,0.82}
\definecolor{backgroundColour}{rgb}{0.95,0.95,0.92}

\lstdefinestyle{CStyle}{
    backgroundcolor=\color{white},   
    commentstyle=\color{mGreen},
    keywordstyle=\color{magenta},
    numberstyle=\tiny\color{mGray},
    stringstyle=\color{mPurple},
    basicstyle=\tiny\ttfamily,
    breakatwhitespace=false,         
    breaklines=true,                 
    captionpos=b,                    
    keepspaces=true,                 
    numbers=left,                    
    numbersep=5pt,                  
    showspaces=false,                
    showstringspaces=false,
    showtabs=false,                  
    tabsize=1,
    language=C
}

\usepackage{algorithm}
\usepackage{algpseudocode}

\usepackage{xspace}
\usepackage{blindtext}
\usepackage{hyperref}

\usepackage{multirow}

\newcommand{\APPR}{\emph{MOTIF}\xspace}
\newcommand{\SEMU}{\emph{SEMu}\xspace}
\newcommand{\SEMUp}{\emph{SEMuP}\xspace}
\newcommand{\MOTIF}{\APPR}

\newcommand{\UTIL}{LibU\xspace}
\newcommand{\ASNLib}{ASNLib\xspace}

\newcommand{\MLFS}{\emph{MLFS}\xspace}

\usepackage{balance}

%% file: authors-IEEE.tex




\author{

\IEEEauthorblockN{
Jaekwon Lee
}
\IEEEauthorblockA{
    \textit{University of Ottawa} \\
    Ottawa, CA \\
    jaekwon.lee@uottawa.ca
}
\and
\IEEEauthorblockN{
Enrico Vigan\`{o}
}
\IEEEauthorblockA{
    \textit{University of Luxembourg} \\
    Luxembourg, Luxembourg \\
    enrico.vigano@uni.lu
}
\and
\IEEEauthorblockN{
Fabrizio Pastore
}
\IEEEauthorblockA{
    \textit{University of Luxembourg} \\
    Luxembourg, Luxembourg \\
    fabrizio.pastore@uni.lu
}
\and
\IEEEauthorblockN{
Lionel Briand
}
\IEEEauthorblockA{
    \textit{University of Ottawa}, Ottawa, CA \\
    \textit{Lero, University of Limerick}, Ireland \\
    lbriand@uottawa.ca
}

}

%% file: sections/abstract.tex
\begin{abstract}
Mutation testing consists of generating test cases that detect faults injected into software (generating mutants) which its original test suite could not.
By running such an augmented set of test cases, it may discover actual faults that may have gone unnoticed with the original test suite.
It is thus a desired practice for embedded software running in safety-critical cyber-physical systems (CPS). 
Unfortunately, the state-of-the-art tool targeting C, a typical language for CPS software, relies on symbolic execution, whose limitations often prevent its application.
\MOTIF overcomes such limitations by leveraging grey-box fuzzing tools to generate unit test cases in C that detect injected faults in mutants. 
Indeed, fuzzing tools automatically generate inputs by exercising the compiled version of the software under test guided by coverage feedback, thus overcoming the limitations of symbolic execution. 
%
Our empirical assessment has shown that it detects more faults than symbolic execution (i.e., up to 47 percentage points), when the latter is applicable.

\end{abstract}

%% file: sections/introduction.tex
\section{Introduction}
\label{sec:introduction}


Mutation testing consists of generating test cases capable of detecting faults not identified by the test suite of the software under test (SUT). Such faults are automatically injected into the SUT; the faulty software versions are called mutants. The proportion of mutants killed (i.e., detected) by a test suite is called mutation score. It has been demonstrated that there exists a strong association between a high mutation score and a high fault revealing capability for test suites~\cite{papadakis2018mutation}. For such reason, mutation testing might be an effective instrument to strengthen test suites for CPS software.

The state-of-the-art (SOTA) mutation testing tool for C (i.e., \SEMU~\cite{Chekam2021}) is based on the KLEE symbolic execution (SE) engine~\cite{KLEE}. Though it is effective with command line utilities, it inherits the limitations of SE. Specifically, it cannot deal with programs requiring complex analyses for input generation (e.g., programs with floating point instructions).

In this paper, we present \MOTIF (MutatiOn Testing wIth Fuzzing), which automates our approach presented at ASE'23~\cite{MOTIF}.
It relies on a grey-box fuzzer, AFL++~\cite{AFL++}, that leverages coverage feedback to learn how to reach deeper into the program without building on heavyweight program analysis or constraint solving. 
The fuzzer provides modified input data, given seed inputs, using the collected code coverage and an evolutionary heuristic approach. 
These characteristics enable us to alleviate the limitations of SE.
\MOTIF implements a pipeline for the fuzzer by automatically generating a test driver that invokes the original and mutated functions with the given fuzzed input; the differential outputs of the functions enables the fuzzer to determine whether inputs kill the mutant.

\MOTIF is available online~\cite{MOTIFGIT}; also, we provide a \emph{replication package} for our experiments~\cite{REPLICABILITY} and a demo video (available at 
\url{https://www.youtube.com/watch?v=peXB6oDOcMY}).

%% file: sections/approach.tex
\section{Tool overview}
\label{sec:approach}

\label{sec:motif}
\begin{figure}[tb]
\begin{center}
\includegraphics[width=8.4cm]{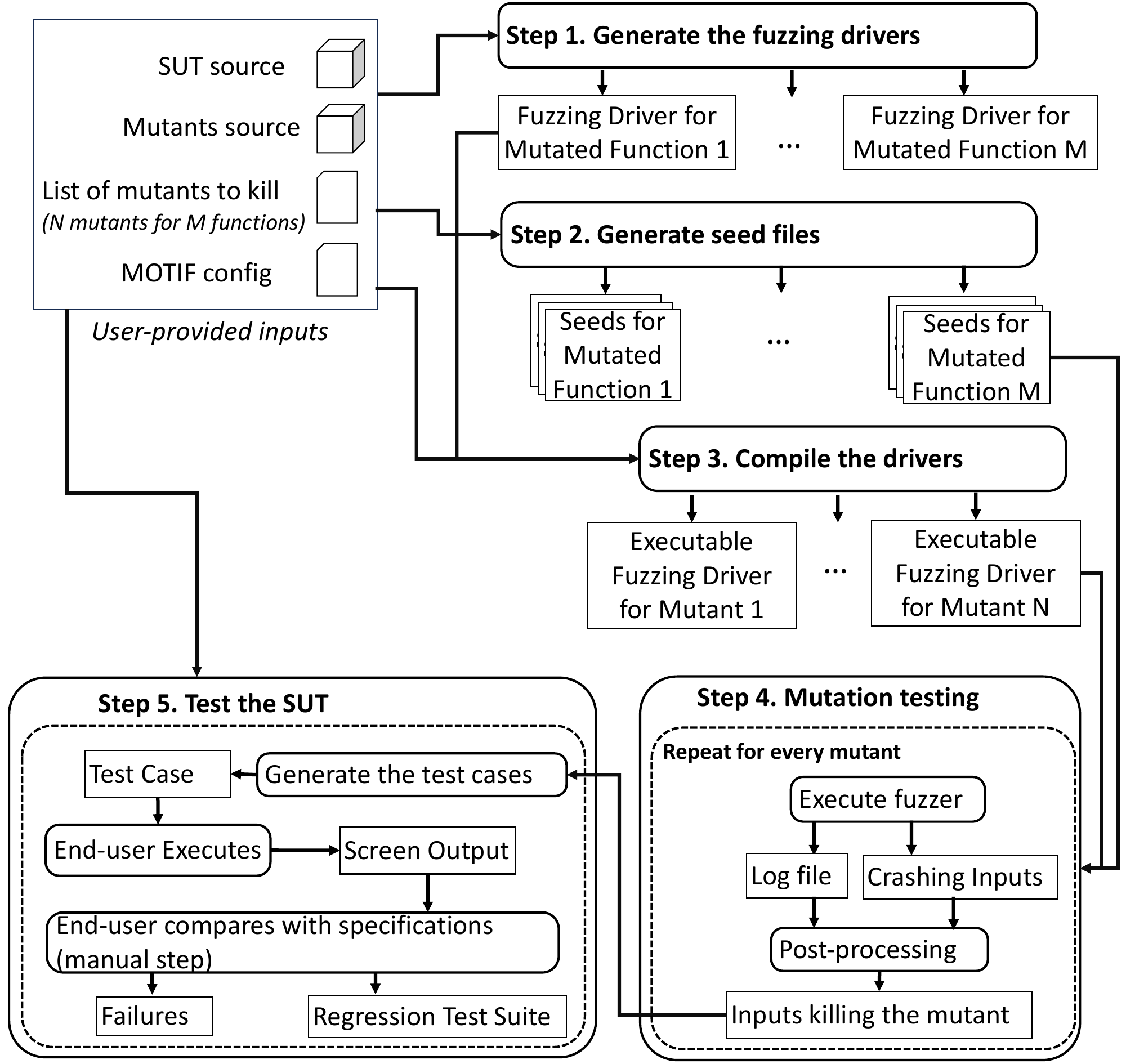}
\vspace*{-0.5em}
\caption{The \MOTIF process.}
\vspace*{-1.8em}
\label{fig:motif:process}
\end{center}
\end{figure}

\MOTIF's workflow is shown in Figure~\ref{fig:motif:process}.
\MOTIF is started by running its Python entry point in a directory selected by the end-user as a workspace. The workspace should include all the inputs required by \MOTIF, which are the SUT source code, the mutants source code, a configuration file for \MOTIF, and a text file with the names of the mutants to kill.
We rely on the mutants generated by MASS~\cite{Oscar:TSE,MASSTOOL} but any mutation analysis tool can be used.

\input{listings/driver.tex}

In Step 1, \MOTIF relies on the \emph{clang} static analysis library~\cite{CLANG} to build an abstract syntax tree of the SUT and determine the types of parameters required by each function under test (the original function the mutant modified). Such information is used to generate a test driver for mutation testing (fuzzing driver); an example is shown in Figure~\ref{asn_driver}. The fuzzing driver has four responsibilities: (1) loading input data that is provided by the fuzzer (Line 2), (2) declaring a set of variables to be used for the parameters of the original and mutated functions (Lines 4-5 and Lines 7-8), (3) invoking the original and mutated functions with the same parameter values filled with the input data (Lines 13-16 and Lines 18-22), and (4) comparing the outputs of both functions (Lines 24-27). Since in C, thanks to pointers and reference arguments, every parameter may be used to store outputs, we compare all the parameters and return values of the original and mutated functions. When differences are observed, the fuzzing driver stops its execution with an abort signal thus letting the fuzzer detect the aborted execution and store the input file (Line 31).
\MOTIF also generates additional test drivers that help with the workflow of \MOTIF.







In Step 2, \MOTIF generates seed files based on the types of input parameters for the function under test. The generated files contain enough bytes to fill all the input parameters with values covering basic cases. Precisely, for each primitive type, we have identified three seed values that are representative of typical input partitions~\cite{MOTIF}. For example, for numeric values, we provide zero, a negative, and a positive number. For each fuzzing driver, \MOTIF generates at most three seed files in such a way that every parameter of the function under test is assigned with each seed value at least once.


In Step 3, \MOTIF compiles the fuzzing driver, the mutated function, and the SUT using the fuzzer compiler; compile commands can be specified in \MOTIF's configuration.  


In Step 4, \MOTIF runs the fuzzer. The execution leads to the generation of fuzzing driver logs and crashing inputs.
\MOTIF processes the corresponding logs to identify likely killed mutants, which happens when either the execution aborts because the generated outputs differ for the original and the mutated functions, or there is a crash in the execution of the mutated function.
Inputs leading to apparently killed mutants are further processed to determine if the function under test generates non-deterministic outputs: \MOTIF executes them with an additional test driver, which is automatically generated and executes the original function twice. If outputs differ, then the function under test is non-deterministic, and the mutant is not deemed killed.

In Step 5, \MOTIF generates a unit test case that is similar to the fuzzing driver.
The generated test case declares a set of arrays initialized with data taken from the input file that killed the mutant; it also declares a set of arrays for the expected outcomes of the original function, which was observed during the mutation testing post-processing step (Step 4). The test case verifies the execution result of the original function with the obtained expected values. 
If the original function is not faulty, the data observed after the execution of the original function can be used in test assertions for regression testing. 

%% file: listings/driver.tex

\begin{figure}
\label{asn_driver} 
\begin{lstlisting}[style=CStyle, tabsize=2, mathescape=true]
int main(int argc, char** argv){
  load_file(argv[1]);  // load the input file and extends it if needed
  /* Variables for the original function */
  T_POS origin_pVal;   // for the first parameter
  int origin_pErrCode; // for the second parameter
  /* Variables for the mutated function */
  T_POS mut_pVal;      // for the first parameter
  int mut_pErrCode;    // for the second parameter
  /* Variables for the return values */
  flag origin_return;  // for the original 
  flag mut_return;     // for the mutant
  /* Copy the input data to the variables for the original function */
  get_value(&origin_pVal, sizeof(origin_pVal), 0); 
  get_value(&origin_pErrCode,sizeof(origin_pErrCode),0); 
  log("Calling the original function");
  origin_return = T_POS_IsConstraintValid(&origin_pVal, &origin_pErrCode);
  /* Copy the same input data to the variables for the mutated function */
  seek_data_index(0); //reset the input data pointer
  get_value(&mut_pVal, sizeof(mut_pVal), 0); 
  get_value(&mut_pErrCode, sizeof(mut_pErrCode), 0); 
  log("Calling the mutated function");
  mut_return = mut_T_POS_IsConstraintValid(&mut_pVal, &mut_pErrCode);

  log("Comparing result values: ");
  ret += compare_value(&origin_pVal, &mut_pVal, sizeof(origin_pVal));
  ret += compare_value(&origin_pErrCode,&mut_pErrCode, sizeof(origin_pErrCode));
  ret += compare_value(&origin_return, &mut_return, sizeof(origin_return));

  if (ret != 0){
    log("Mutant killed");
    safe_abort();
  }
  log("Mutant alive");
  return 0;
}    
\end{lstlisting}
\vspace{-0.8em}
\caption{Example fuzzing driver for the \ASNLib{} subject.}
\vspace{-1em}
\end{figure}

%% file: sections/evaluation.tex
\section{Empirical Evaluation}
\label{sec:evaluation}

We performed experiments with software used in commercial satellites: 
(a) \MLFS{}, the Mathematical Library for Flight Software~\cite{MLFS}, (b) \UTIL{}, which is a utility library developed by one of our industry partners, and (c) \ASNLib{}, a serialization/deserialization library.
We used \MOTIF to kill the mutants generated by MASS and not killed by the original test suite.
We compared \MOTIF with \SEMU by creating a modified version of \MOTIF (hereafter, \SEMUp) that relies on \SEMU to kill mutants. We applied the two tools on subjects processable by KLEE. We executed each tool ten times for each subject, for 10,000 seconds. 

Results show that  
\MOTIF outperforms \SEMUp; indeed, across the ten runs, it kills a percentage of mutants that is 46.86 percentage points (pp) and 10.52 pp higher than \SEMUp's, for \UTIL{} and \ASNLib{}, respectively. The difference is significant at every timestamp, based on Fisher's test.

We also applied \MOTIF on subjects that cannot be tested with \SEMUp: all the mutants of \MLFS (3,891) and a subset of the \UTIL{} mutants (290). 
For \MLFS and \UTIL{}, 
\MOTIF kills on average 1,399.4 (35.97\%) mutants and  120 (41.38\%) mutants, respectively.
Percentages are lower than the ones presented above because the nature of these two subjects (e.g., mathematical functions) leads to mutants killed only by inputs from a very small part of the input domain.

%% file: sections/conclusion.tex
\section{Conclusion}
\label{sec:conclusion}

\MOTIF is a tool leveraging grey-box fuzzing to perform mutation testing at the unit level. 
Our empirical results have shown that it outperforms a SOTA approach based on symbolic executions and detects a large number of mutants 
for subjects where symbolic execution is infeasible. 
Our results imply that leveraging fuzzing tools can advance mutation testing. 
For future work, we will extend \MOTIF to C++. We will also investigate the effectiveness improvements obtained with different fuzzing configurations and tools.   


%% file: MAIN.bbl
\begin{thebibliography}{10}
\providecommand{\url}[1]{#1}
\csname url@samestyle\endcsname
\providecommand{\newblock}{\relax}
\providecommand{\bibinfo}[2]{#2}
\providecommand{\BIBentrySTDinterwordspacing}{\spaceskip=0pt\relax}
\providecommand{\BIBentryALTinterwordstretchfactor}{4}
\providecommand{\BIBentryALTinterwordspacing}{\spaceskip=\fontdimen2\font plus
\BIBentryALTinterwordstretchfactor\fontdimen3\font minus \fontdimen4\font\relax}
\providecommand{\BIBforeignlanguage}[2]{{%
\expandafter\ifx\csname l@#1\endcsname\relax
\typeout{** WARNING: IEEEtran.bst: No hyphenation pattern has been}%
\typeout{** loaded for the language `#1'. Using the pattern for}%
\typeout{** the default language instead.}%
\else
\language=\csname l@#1\endcsname
\fi
#2}}
\providecommand{\BIBdecl}{\relax}
\BIBdecl

\bibitem{papadakis2018mutation}
M.~Papadakis, D.~Shin, S.~Yoo, and D.-H. Bae, ``{Are mutation scores correlated with real fault detection? A large scale empirical study on the relationship between mutants and real faults},'' in \emph{2018 IEEE/ACM 40th International Conference on Software Engineering (ICSE)}.\hskip 1em plus 0.5em minus 0.4em\relax IEEE, 2018, pp. 537--548.

\bibitem{Chekam2021}
T.~T. Chekam, M.~Papadakis, M.~Cordy, and Y.~L. Traon, ``Killing stubborn mutants with symbolic execution,'' \emph{ACM Transactions on Software Engineering and Methodology}, vol.~30, no.~2, Jan. 2021.

\bibitem{KLEE}
C.~Cadar, D.~Dunbar, and D.~Engler, ``Klee: Unassisted and automatic generation of high-coverage tests for complex systems programs,'' in \emph{Proceedings of the 8th USENIX Conference on Operating Systems Design and Implementation}, ser. OSDI'08, vol.~8.\hskip 1em plus 0.5em minus 0.4em\relax USA: USENIX Association, 2008, p. 209–224.

\bibitem{MOTIF}
\BIBentryALTinterwordspacing
J.~Lee, E.~Viganò, O.~Cornejo, F.~Pastore, and L.~Briand, ``Fuzzing for cps mutation testing,'' in \emph{Proceedings of the 38th IEEE/ACM International Conference on Automated Software Engineering (ASE 2023)}, 2023. [Online]. Available: \url{https://arxiv.org/abs/2308.07949}
\BIBentrySTDinterwordspacing

\bibitem{AFL++}
A.~Fioraldi, D.~Maier, H.~Ei{\ss}feldt, and M.~Heuse, ``{AFL++} : Combining incremental steps of fuzzing research,'' in \emph{14th USENIX Workshop on Offensive Technologies (WOOT 20)}.\hskip 1em plus 0.5em minus 0.4em\relax USENIX Association, Aug. 2020.

\bibitem{MOTIFGIT}
J.~Lee, E.~Viganò, O.~Cornejo, F.~Pastore, and L.~Briand, ``{MOTIF toolset},'' \url{https://github.com/SNTSVV/MOTIF}, 2023.

\bibitem{REPLICABILITY}
------, ``{Replication package},'' \url{https://doi.org/10.6084/m9.figshare.22693525}, 2023.

\bibitem{Oscar:TSE}
\BIBentryALTinterwordspacing
O.~E. Cornejo~Olivares, F.~Pastore, and L.~Briand, ``{Mutation Analysis for Cyber-Physical Systems: Scalable Solutions and Results in the Space Domain},'' \emph{IEEE Transactions on Software Engineering}, vol.~48, no.~10, pp. 3913--3939, 2022. [Online]. Available: \url{https://doi.org/10.1109/TSE.2021.3107680}
\BIBentrySTDinterwordspacing

\bibitem{MASSTOOL}
O.~Cornejo, F.~Pastore, and L.~Briand, ``Mass: A tool for mutation analysis of space cps,'' in \emph{2022 IEEE/ACM 44th International Conference on Software Engineering: Companion Proceedings (ICSE-Companion)}, 2022, pp. 134--138.

\bibitem{CLANG}
{LLVM project}, ``{Clang} library,'' https://clang.llvm.org/, 2023.

\bibitem{MLFS}
\BIBentryALTinterwordspacing
{European Space Agency}, ``{MLFS} - mathematical library for space software,'' 2021. [Online]. Available: \url{https://essr.esa.int/project/mlfs-mathematical-library-for-flight-software}
\BIBentrySTDinterwordspacing

\end{thebibliography}
